\documentclass[aps, prb, amsmath, twocolumn, showpacs, reprint, superscriptaddress]{revtex4-1}
\usepackage{graphicx}
\usepackage{color}

\begin{document}

% AUTHORS
%========

\author{Peter Kroiss}
\affiliation{Department of Physics, Arnold Sommerfeld Center for Theoretical Physics and Center for NanoScience, Ludwig-Maximilians-Universit\"at M\"unchen, Theresienstrasse 37, 80333 Munich, Germany}
\author{Massimo Boninsegni}
\affiliation{Department of Physics, University of Alberta, Edmonton, Alberta, Canada}
\author{Lode Pollet}
\affiliation{Department of Physics, Arnold Sommerfeld Center for Theoretical Physics and Center for NanoScience, Ludwig-Maximilians-Universit\"at M\"unchen, Theresienstrasse 37, 80333 Munich, Germany}

% Title and Abstract
%===================

\title{Ground state phase diagram of  Gaussian-Core bosons in two dimensions}
\date{\today}
\pacs{03.65.Vf, 67.10.-j}

\begin{abstract}
The ground state 
of a two-dimensional (2D) system of Bose particles 
of spin zero,
interacting via a repulsive Gaussian-Core potential, has been investigated by means of Quantum Monte Carlo simulations.  The quantum phase diagram is 
qualitatively identical to that of 2D Yukawa bosons. While the system is a fluid at all densities for weak coupling, 
in the strong coupling regime it transitions upon compression from a low density superfluid to a crystal,  and then into a reentrant 
superfluid phase.
No evidence of a (supersolid) cluster crystal phase  is seen.

\end{abstract}

\maketitle

% Introduction
%=============
\section{Introduction}

In the absence of disorder, frustration, or an external potential  the ground state of an interacting scalar Bose system is always ordered, {\it i.e.}, a 
well-defined symmetry of the Hamiltonian is spontaneously broken. 
Only two types of order are believed to be possible, namely crystalline order, which breaks translational symmetry, or off-diagonal long-range order (superfluid), 
in which case the global $U(1)$ symmetry is broken. \\ \indent
Normally, only one of these two types of order is present,  as a result of the competition between particle interactions, typically favoring crystallization, and 
quantum delocalization, promoting superfluidity. 
Such is the case for $^4$He, featuring either a superfluid or an insulating crystal in the ground state, depending on the external pressure.\cite{ceperley1995} No exceptions 
are known in continuous space when the interaction potential is of Lennard--Jones type.
However, {\em supersolid} ground states, simultaneously displaying both types of order,\cite{boninsegni2012} have been predicted for a class of Bose systems with 
pair-wise interparticle potentials 
featuring a soft and flat repulsive core at short distances. The supersolid phase arises through the formation at high density of a cluster crystal (CC) 
with more than one particle per unit cell.
At sufficiently low temperature, particle tunnelling across adjacent clusters  establishes superfluid  phase coherence throughout the whole system.\cite{cinti2010,saccani2011, boninsegni2012B}
\\ \indent
Cluster crystals have been extensively investigated in the context of classical soft-core systems.\cite{likos2001}
It has been conjectured \cite{rossi2011} that a necessary condition for the presence of a CC phase in a soft-core system is that the Fourier transform of the potential go negative in 
a wave vector range close to $k \sim 1/d$, with $d$ the range of the soft core.
Computer simulations of a classical two-dimensional (2D) system of particles interacting through a Gaussian-Core\cite{stillinger1976} potential
\begin{equation}\label{pot}
V(r)=\epsilon\ {\rm exp}\biggl[ -\frac{r^2}{2\sigma^2}\biggr ]
\end{equation}
whose Fourier transform is positive-definite, has yielded no evidence of a CC at low temperature,~\cite{prestipino2011}   thus supporting the hypothesis of
Ref. \onlinecite{rossi2011}.
The classical  ground state is a crystal at all densities; at low temperature,  equilibrium low- and high-density  fluid phases exist on both sides of the crystal, 
with  hexatic phases, characterized by the absence of positional order but by a non-vanishing orientational order parameter, separating the crystal from the fluid.
\\ \indent
An interesting theoretical question is to what a degree quantum-mechanical effects alter the classical phase diagram.  Mean field theoretical treatments based on 
the Gross-Pitaevskii\cite{GP} equation, suggest that a negative Fourier component in the pair potential 
is a necessary condition for a roton instability toward crystallization.\cite{roton,roton2,roton3,roton4} On the other hand, such an approach essentially describes 
a supersolid as a superfluid with a density modulation, and is therefore applicable to crystals with a very large number of particles per unit cell. If the number 
of particles per unit cell is only a few,  it is known that  Bose statistics can considerably extend the domain of existence of the CC in soft-core systems, with 
respect to what one would observe classically.\cite{cinti2014}  Thus,  it is conceivable that a CC phase (turning superfluid at low temperature)  could be stabilized 
by quantum-mechanical exchanges in a system of Gaussian-Core bosons. Also, 
the investigation of a quantum-mechanical version of the Gaussian-Core model can offer insight into the role of those quantum fluctuations in the context of soft matter systems.\cite{note0}
\\ \indent
In this work, we present results of extensive Quantum Monte Carlo simulations at low temperature of a 2D system of  spin-zero particles interacting via a Gaussian pair potential. 
Our study shows that, as expected, quantum effects strengthen the fluid phase, which extends all the way to temperature $T=0$ in a wide region of the quantum phase diagram. 
No cluster crystal and no supersolid phase is found. Indeed, superfluid and (insulating) 
crystal are the only two phases observed. The resulting quantum phase diagram is qualitatively identical with that of 2D Yukawa bosons,\cite{Magro93, Osychenko2012} suggesting  that it may 
generically describe all Bose systems featuring the same type of repulsive interaction at short distances, i.e., one that is strong enough to prevent the formation of clusters 
of particles, but not enough to stabilize the crystalline phase at high density.
\\ \indent
The remainder of this paper is organized as follows: in Sec.~\ref{modset} we describe the model of our system of interest and provide details of the calculation; in Sec.~\ref{res} 
we illustrate our results and outline our conclusions in Sec.~\ref{conc}.

% Theoretical setup
%==================
\section{Model and Methodology}\label{modset}
We consider a system of $N$ spin-zero bosons of mass $m$, enclosed in a simulation box with periodic boundary conditions in both directions. 
The aspect ratio of the box is designed to fit a triangular solid.
Particles interact via the pair 
potential described by Eq.~\ref{pot}.
The many-body Hamiltonian of the system in reduced units reads
\begin{equation}\label{ham}
H =- \frac{\Lambda}{2} \sum_{i=1}^N \nabla_i^2 + \sum_{i<j} {\rm exp}\biggl [-\frac {1}{2}\ r_{ij}^2\biggr ],
\end{equation}
where ${\bf r}_i$ is the position vector of the $i$th boson, $r_{ij}\equiv |{\bf r}_i-{\bf r}_j|$, and $\Lambda\equiv \hbar^2/(m\epsilon\sigma^2)$ is the quantum coupling constant.
All lengths are expressed in units of $\sigma$, whereas $\epsilon$ sets the energy and temperature scale (Boltzmann's constant $k_B$ is set to one). 
Besides $\Lambda$, the only other parameter of the system at  temperature $T=0$  is the density $\rho$, or, equivalently, the (dimensionless) mean interparticle distance $r_s=1/\sqrt{\rho\sigma^2}$.
\\ \indent
We obtained the thermodynamic phase diagram  by means of Quantum Monte Carlo
simulations based on the Worm Algorithm in the continuous-space path integral representation.
Because this well-established computational methodology is thoroughly described elsewhere,~\cite{bps2006a, bps2006b} we do not review it here. The most important aspect to be emphasized 
 is that it enables one to compute thermodynamic properties of Bose systems at finite temperature, directly from the microscopic Hamiltonian. In particular, energetic, structural 
and superfluid properties can be computed in practice with no approximation. 
\\ \indent
Technical aspects of the calculations are standard; we carried out simulations of systems comprising up to 1024
particles, using the standard form\cite{bps2006b} for the 
high-temperature  imaginary-time propagator accurate up to fourth order in the time step $\tau$; the smoothness of the potential, however, allows one to obtain accurate results with 
the primitive form as well, with comparable CPU time. We identify the different thermodynamic phases 
(superfluid and crystalline) through the computation of the superfluid density, using the well-known winding number estimator,
\cite{pollock1987} the one-body density matrix, as well as the pair correlation function. As we aim at obtaining the ground state ($T=0$) phase diagram, we performed calculations 
at temperatures sufficiently low not to see any changes in the values of cogent physical quantities (e.g., the energy) within the statistical uncertainties of the calculations; 
typically, this means $T\lesssim T^\star\equiv\Lambda/ r_s^{2}$.

% Discussion of the phase diagram
%================================

%------------------------------------------------------------
\begin{figure}[ptb]
\includegraphics[width=\linewidth]{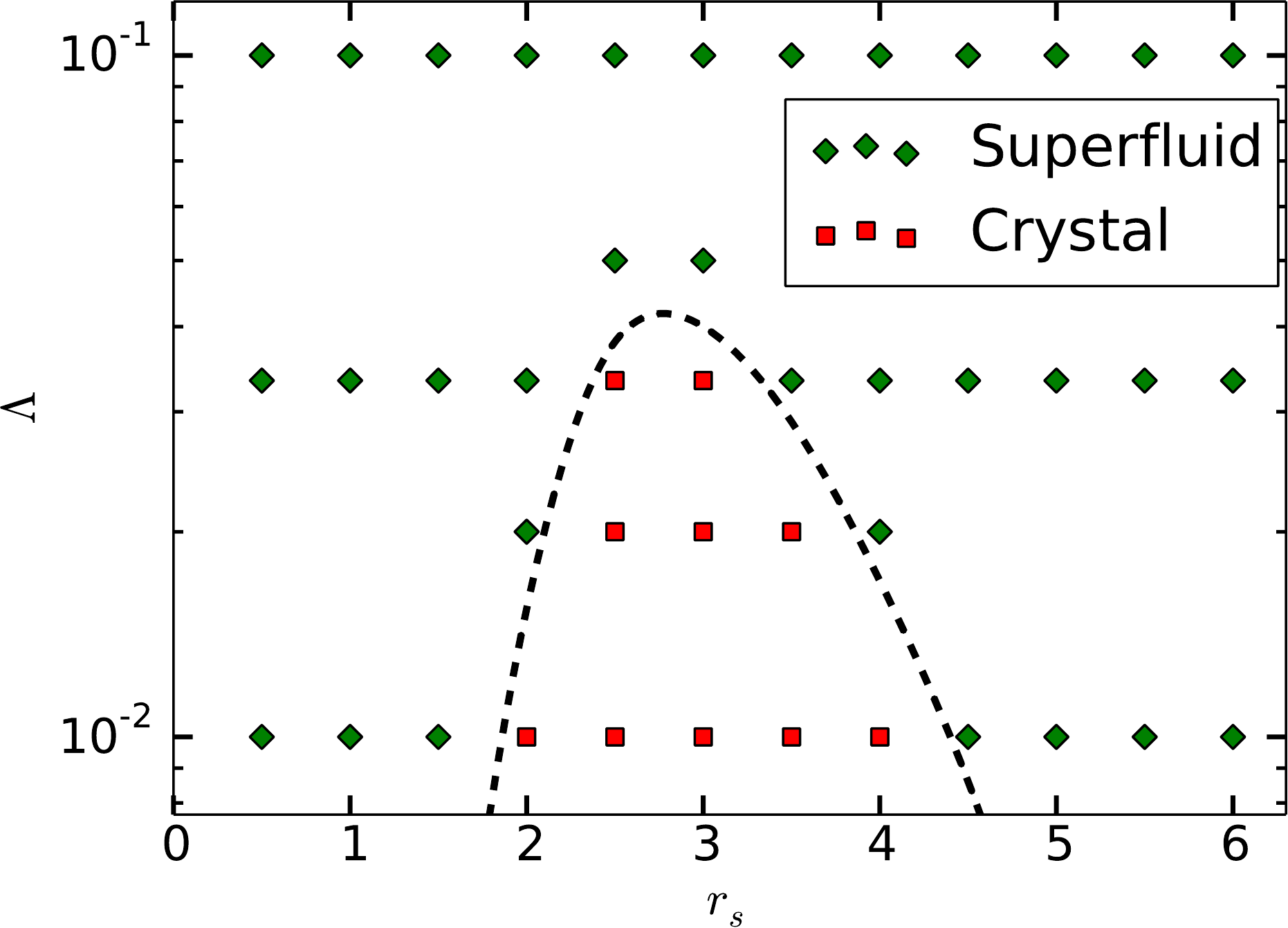}
\caption{\label{fig:phase_diagram}  
(Color online).  Ground state phase diagram of a Bose system with Gaussian-Core interaction for different values of the parameter $\Lambda$ (see text) and mean interparticle spacing $r_s$. 
For low values of $\Lambda$ a crystal phase becomes increasingly stable. At low densities a superfluid phase is seen, while a reentrant superfluid
phase is found for high densities. Numerical data are represented by symbols, the dashed line is a guide for the eye.
}
\end{figure}
%------------------------------------------------------------

\section{results}\label{res}
\subsection{Phase Diagram}

Our findings are summarized in Fig.~\ref{fig:phase_diagram}, showing the ground state phase diagram of the system, as described by the Hamiltonian (\ref{ham}), in the $(r_s$--$\Lambda)$ plane.
All of the results presented here are extrapolated to the $\tau\to 0$ limit.
We identify the following phases:
(i) a superfluid phase at all densities for $\Lambda \gtrsim 0.03$, and in the low and high density limit for lower values of $\Lambda$, where the physics of the system 
is dominated by quantum delocalization and Bose statistics. All of our numerical data at finite temperature show consistency with the Berezinsky-Kosterlitz-Thouless (BKT) scenario of the superfluid 
transition in 2D,\cite{Berenzinsky1971, KT} {\it i.e.}, 
the transition between a quasi-long-range superfluid and a normal phase at finite temperature occurs through the standard unbinding of vortex--anti-vortex pairs with charge 1.
Unconventional are however the microscopic properties of the reentrant superfluid phase (see below).
(ii) A crystalline (triangular) phase becomes stable for $\Lambda\lesssim 0.03$, centered around $r_s = 3$. It extends its domain of existence as $\Lambda \to 0$, as 
the potential energy plays an increasingly important role. 
We found no evidence of other phases, such as supersolid phases.

On general grounds we expect a superfluid phase  at $T=0$ in the low density limit by analogy to superfluid helium: In the dilute limit, the potential energy is much smaller than the 
energy of zero point fluctuations, {\it i.e.}, 
the quantum pressure prevents crystallization. That a superfluid phase might also occur at high densities is different from helium: In the present case the soft-core of the potential 
is unable to prevent the overlap of particles 
at high enough density, leading to a reentrant superfluid, just as in the phase diagram of 2D 
Yukawa bosons.\cite{Magro93, Osychenko2012}

%------------------------------------------------------------
\begin{figure}[ptb]
\includegraphics[width=\linewidth]{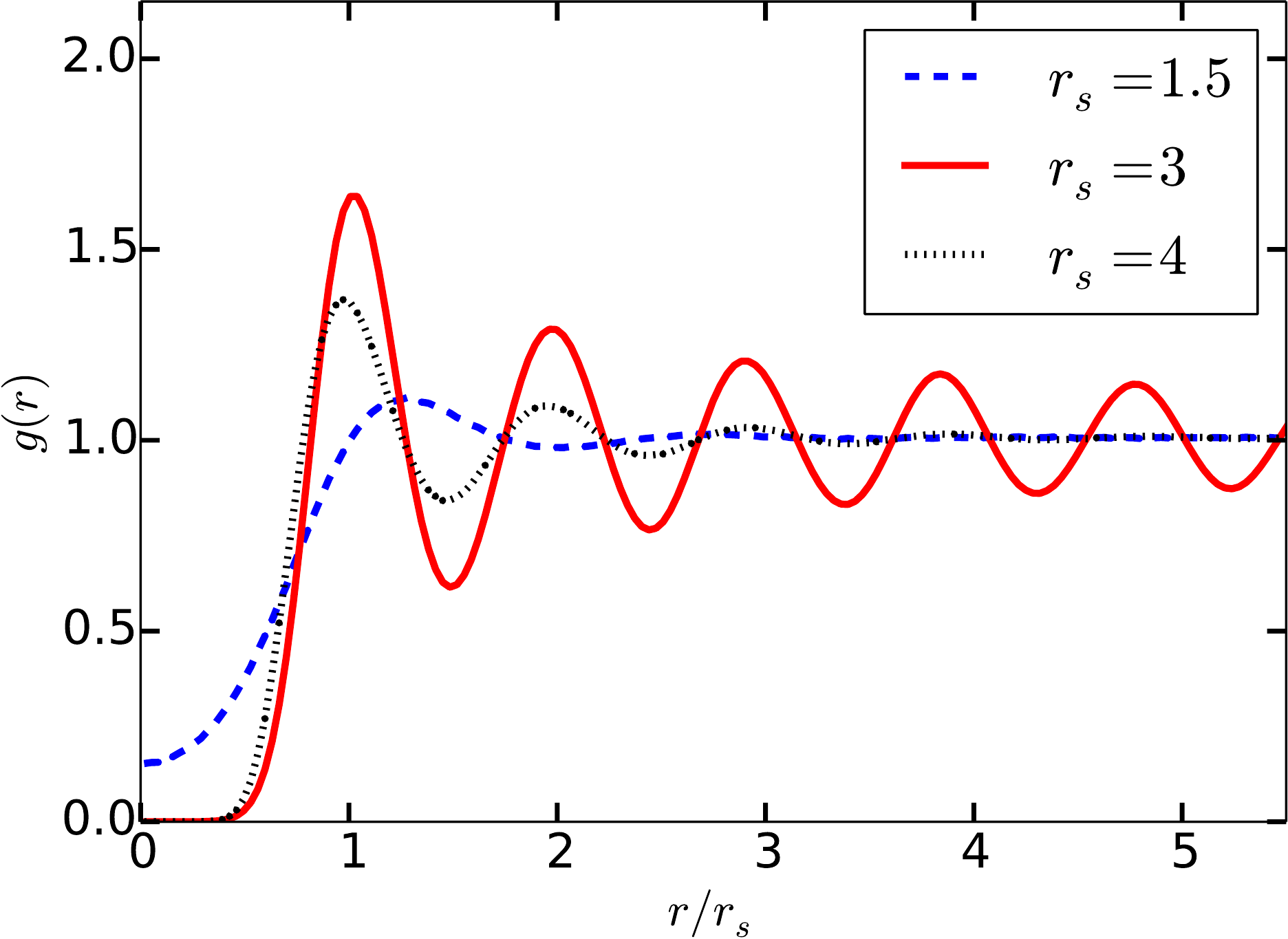}
\caption{\label{fig:correlation_functions} 
(Color online). Ground state pair correlation functions for different values of $r_s$ at $\Lambda=1/30$.
Error bars are too small to be seen on the scale of the figure.
While the pair correlation functions for $r_s = 4$ and $r_s = 3$ (corresponding to the superfluid and crystal phases) show hard-core separation of particles, the reentrant 
superfluid phase ($r_s = 1.5$) acquires a finite value at the origin. In this phase, only very weak peaks are left, rendering $g(r)$ essentially flat for $r \gtrsim 2$.
}
\end{figure}
%------------------------------------------------------------

% Discussion of Correlation functions
%====================================

Insight into the structure of the various phases is offered by the pair correlation function $g(r)$, shown in 
Fig.~\ref{fig:correlation_functions}  for the crystal and both conventional and reentrant superfluids.
For $r_s \ge 3$ ({\it i.e.}, in the crystalline and low density superfluid phases), the physics effectively mimics that of a hard-core system, characterized  by a vanishing $g(r)$ at short distances, 
and resulting in conventionally 
looking pair correlation functions, as is shown in Fig.~\ref{fig:correlation_functions} for $r_s=3$.
The peak structure in $g(r)$ is washed out as the system is compressed and $r_s$ is reduced below 3, at which point $g(r)$ suddenly acquires a finite value at the origin, 
as the finite potential energy cost no longer prevents  particles from overlapping. Further compression of the system into the reentrant superfluid phase has the effect of raising 
the value of $g(0)$, as the system approaches the behavior of a free Bose gas. Note that the first peak of $g(r)$ in the conventional superfluid phase is more pronounced than the
corresponding peak of the reentrant superfluid; this is a consequence of the effective hard-core interaction between the particles.
As a finite-temperature method is employed, the magnitude of the peaks of $g(r)$ decreases even in the solid phase. This is expected, as thermal fluctuations do not allow true crystalline 
order in two dimensions. The important observation is that the distance between neighboring peaks is constant over a large range of $r$.

% Discussion of r_s = 1, Lambda = 1/1000 point
%=============================================

%------------------------------------------------------------
\begin{figure}[ptb]
\includegraphics[width=\linewidth]{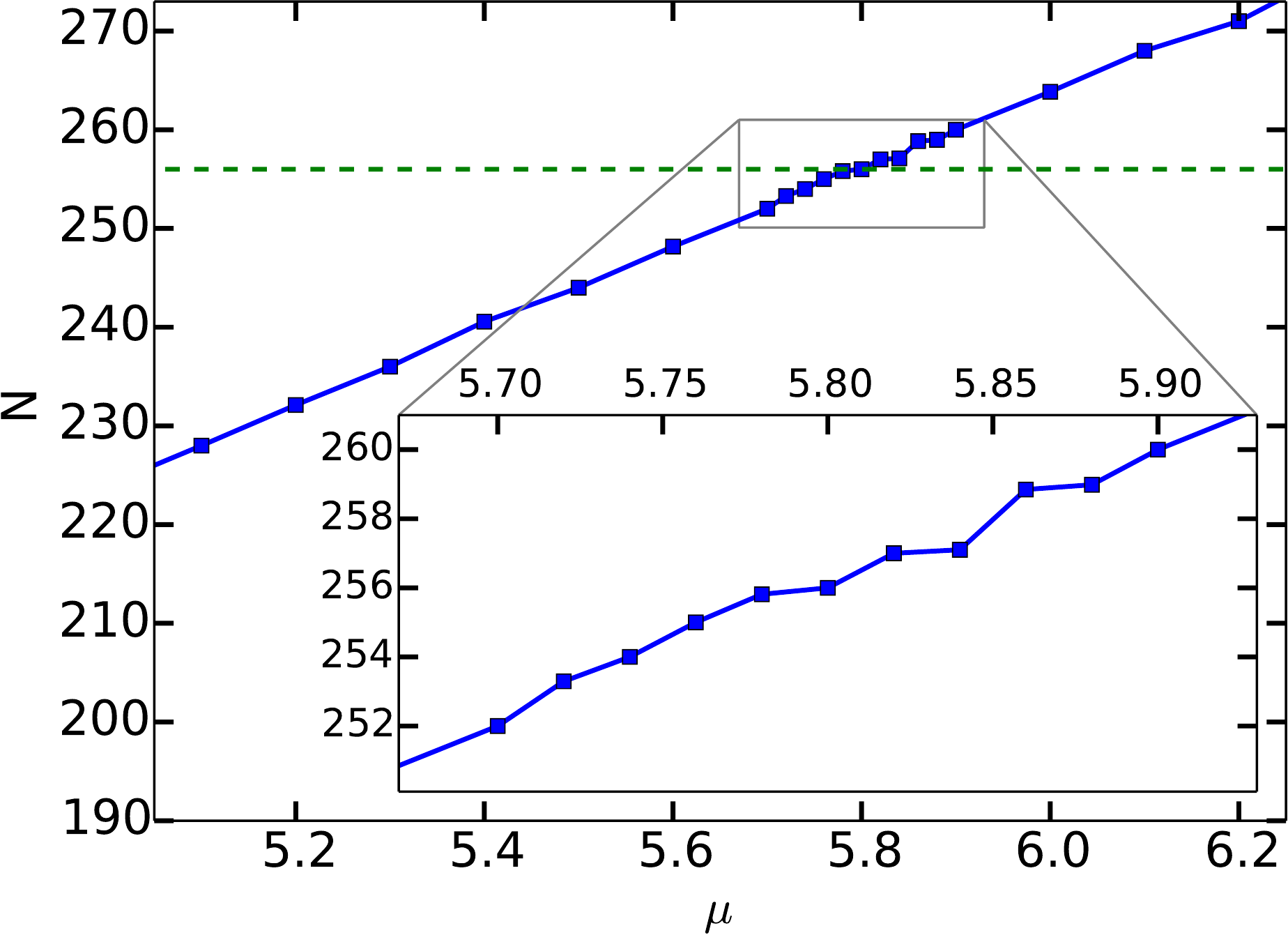}
\caption{\label{fig:nmu} 
(Color online).  Particle number as a function of grand-canonical chemical potential $\mu$ for $r_s = 1$, $T = 0.001$ and $\Lambda = 1/1000$.
The initial configuration is always a solid configuration at $N= 256$ in a commensurate box.
As the grand-canonical simulation allows the particle number to fluctuate, non-integer values of $N$ are possible.
However, for the point of $\mu = 5.8$ (corresponding to $N = 256$), no particle number changes are observed; this happens
for several values of $\mu$. The horizontal line corresponds to 256 particles.
Error bars are not visible on this scale.
In the inset, a zoom on the data shows regions of nearly constant particle number.
}
\end{figure}
%------------------------------------------------------------

%------------------------------------------------------------
\begin{figure}[ptb]
\includegraphics[width=\linewidth]{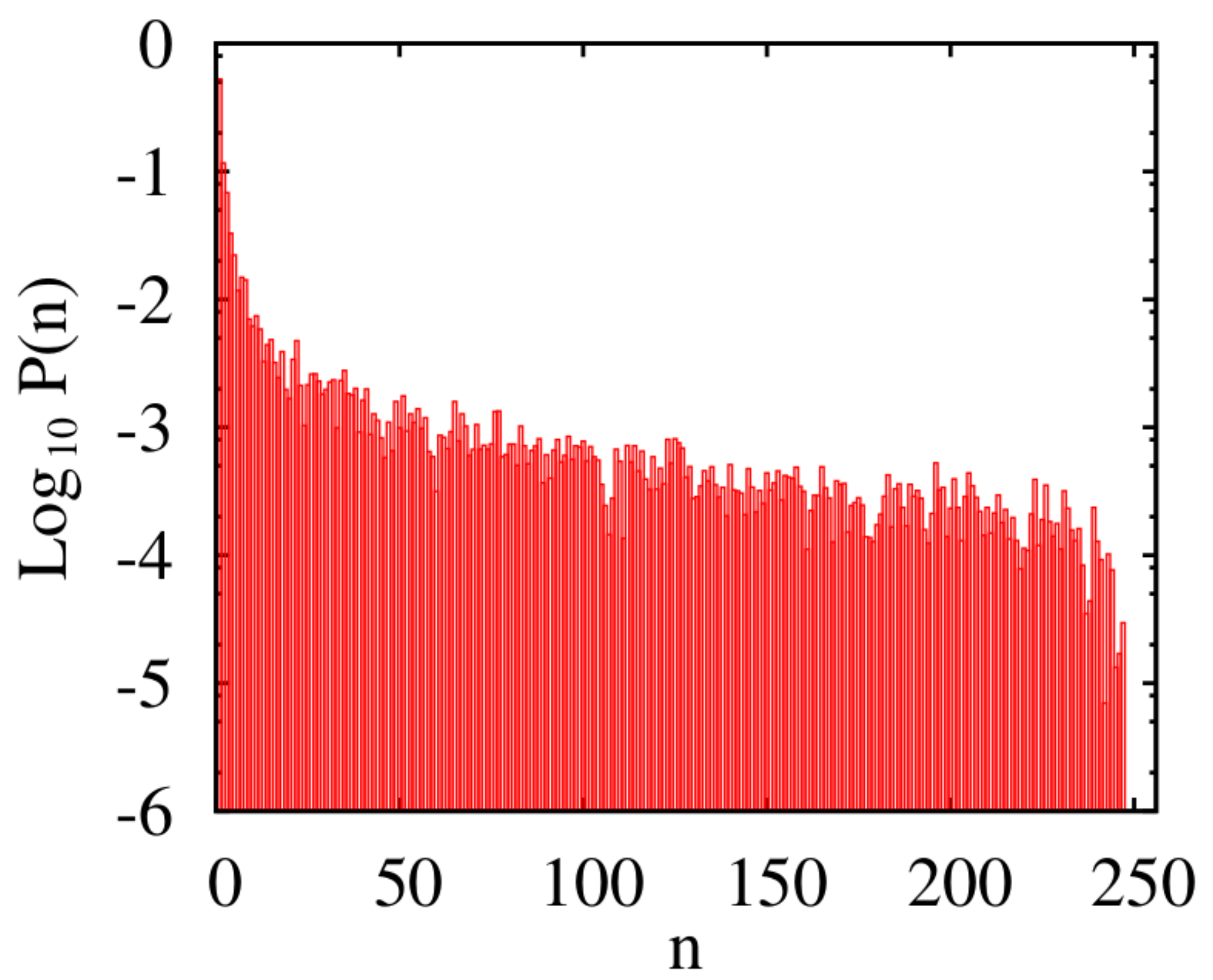}
\caption{\label{fig:cycle} 
(Color online). For $r_s = 1$, $\Lambda = 0.001$, $T = 1/6400$ and 256 particles, the probability $P(n)$ of bosonic exchanges
involving $n$ particle worldlines is shown.
}
\end{figure}
%------------------------------------------------------------

%------------------------------------------------------------
\begin{figure}[ptb]
\includegraphics[width=\linewidth]{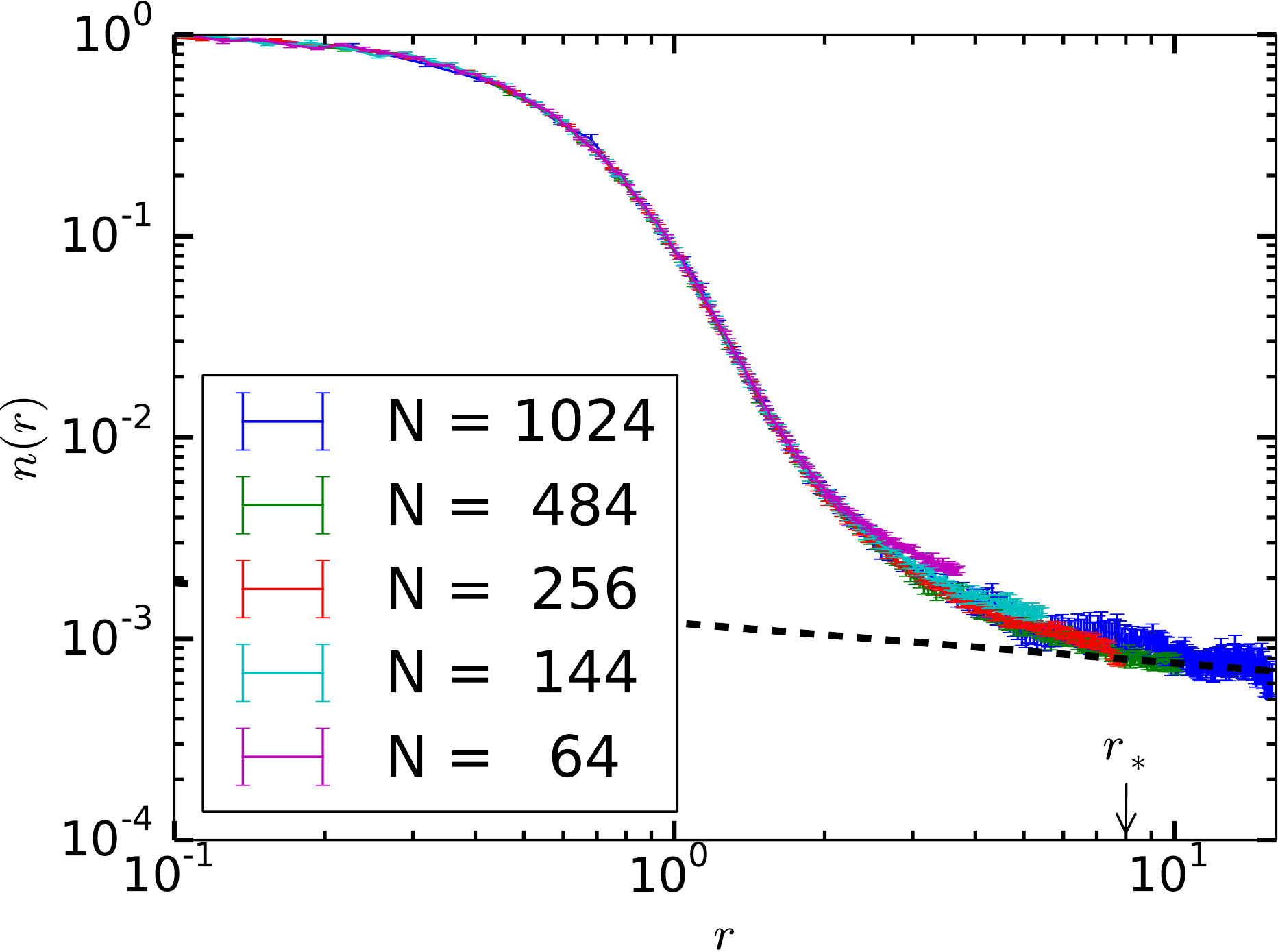}
\caption{\label{fig:obdm} 
(Color online). The equal-time one-body density matrix at $r_s = 1$ for $\Lambda = 0.002$, $T = 0.002$ and different system sizes.
After a rapid initial decay, a non-integrable power-law decay is seen for big enough system sizes, rendering the phase a superfluid. 
The dashed line illustrates the linear regime which holds for all $r > r_*$ (see text).
}
\end{figure}
%------------------------------------------------------------

\subsection{Reentrant superfluid phase in the limit of low $\Lambda$}

Next, we investigate the point $r_s = 1$, $\Lambda = 0.001$ and $T = 0.001$ which corresponds to high density and weak quantum fluctuations.
Fig.~\ref{fig:nmu} shows the total number of particles for different grand-canonical simulations \cite{bps2006b}
in an initially solid configuration of $N = 256$ particles
for several values of the chemical potential $\mu$.
While the curve suggests a linear relationship between chemical potential and particle number 
(implying a constant non-zero compressibility),
tiny deviations of much lower compressibility can be seen.
These may hint at a tendency towards insulating behavior, but this is not the case here: 
They are a consequence of low temperature and finite system size, similar to the observation 
of finite charging levels in a quantum dot.
It is well known that in dilute superfluids the compressibility at very low temperatures can also
be very small on small system sizes and very low temperatures due to the same mechanism. 
What is surprising here, is that this occurs already for temperatures of the order of the Kosterlitz-Thouless temperature. 
Nevertheless, a gapped solid-like structure can certainly be ruled out in the thermodynamic limit
on the basis of the pair correlation function.
In addition, the corresponding Green's function at zero momentum, $G(\tau, p = 0)$, goes up with increasing system size for $|\tau| \gg 0$, {\it i.e.},
adding particles to the simulation becomes easier.
This is one of the manifestations that this parameter regime 
is very difficult to simulate. 
In particular, the superfluid density has anomalously large autocorrelation times, which are unusual for the worm algorithm.
Interestingly, bosonic particle exchanges do not suffer from the same decorrelation problem:
Fig.~\ref{fig:cycle} shows a typical distribution of particle permutations where exchanges up to the total number of particles are reached.
The distribution can already be reliably measured in early stages of the simulation without observing any superfluid response.
This disagrees with the perception that macroscopic exchanges directly trigger superfluidity (which holds for dilute systems).
The superfluidity of the phase is never in doubt though, as can be seen from the the one-body density matrix $n(r)$, shown for a similar point ($\Lambda = 0.002$) in Fig.~\ref{fig:obdm}.
It experiences a weak power-law decay (setting in at a distance $r_*$) after a fast initial drop with a power less than $1/4$, demonstrating the existence of off-diagonal long-range order in the 
system. Although the asymptotic behavior of the curve is consistent with conventional superfluids, the low value of $n(r_*)$ at which this power-law sets in, is unusual. Comparing this 
curve with measurements for higher $\Lambda$, it follows that we can tune $n(r_*)$ with $\Lambda$. 
For increasing $\Lambda$, the winding estimator yields the correct
superfluid response more and more reliably (cf. Fig.~\ref{fig:phase_diagram}).

On the basis of all these observations, we can state that the system
is ultimately a superfluid based on its properties for big enough system sizes.
The unusual microscopics are due to the denseness witnessed in this parameter regime.
Finally, we note that the behavior of $n(r)$ and $g(r)$ for sufficiently large values of $r$, as well as of the superfluid density, is remarkably 
similar to the observations of the superglass in Ref.~\onlinecite{SG}. 
However, as we are looking for the thermodynamic ground state of the system, such a metastable state can be excluded.
Hence, the claim for a superglass, as in Ref.~\onlinecite{SG}, should only be made on the basis of additional real-time considerations.
We leave for future work the static response of this phase, \textit{i.e.},  how it responds to pinning or disorder.

% Conclusion
%===========

\section{Conclusion}\label{conc}

In conclusion, a first principles numerical investigation of the phase diagram of a 2D Bose system with Gaussian-Core pair-wise
interactions has yielded two different phases: A crystal and a superfluid, which also shows reentrant behavior at high densities.
No supersolid or cluster crystal phases were found.
This was anticipated by the positiveness of the interaction potential in Fourier space and affirms the cluster crystal conjecture of Ref.~\onlinecite{rossi2011}.

The reentrant superfluid phase demonstrates unexpected behavior for high particle mass: 
The power-law decay of the one-body density matrix sets in at large distances, where its value is already quite low.
This requires big system sizes to capture the relevant length scales.
Likewise, grand-canonical simulations experience deviations from non-zero compressibility for finite systems.
This is complemented by the occurrence of large cycles of particle wordline permutations, independent of system size.

Whether such a system may lend itself to experimental realization is difficult to assess. Recent progress in cold atom manipulation 
allows one to tailor, to some degree, the interaction among atoms.\cite{zwerger2008} The other aspects of the system, including its detection, are already well within current technology.\cite{zwerger2008} 
The crucial part is that the Gaussian potential has no preferred length scale, unlike the softened dipolar~\cite{moroni} or Rydberg potentials, whereas a system of Yukawa bosons~\cite{Magro93, Osychenko2012} 
shows a qualitatively similarly 
looking phase diagram.

% Acknowledgements and grants
%============================

We would like to thank W.~Krauth, J.~Nespolo, T.~Pfef\-fer, N.~Prokof'ev and B.~Svistunov for fruitful discussions.
This work is supported by the Excellence Cluster NIM, FP7/Marie-Curie Grant No.~321918 (FDIAGMC), FP7/ERC Starting Grant No.~306897 (QUSIMGAS), as well as by the Natural Science and Engineering Research Council of Canada. 
Computing support from Westgrid is gratefully acknowledged.

% Bibliography
%=============

\end{document}